\documentstyle[twocolumn,aps]{revtex}
\begin{document}
\draft
\title{\bf {Temperature enhanced persistent currents and ``$\phi_0/2$
periodicity''}}
\author{M. V. Moskalets$^a$ and P. Singha Deo$^{b,}$\cite{eml}}
\address{$^a$Fl.48, 93-a prospect Il'icha, 310020 Khar'kov, Ukraine\\
      $^b$S.N. Bose National Centre for Basic Sciences, JD Block,
      Sector 3, Salt Lake City, Calcutta 98, India.}
\maketitle
\begin{abstract}
We predict a non-monotonous temperature dependence of the 
persistent currents in a ballistic ring coupled strongly to a
stub in the grand canonical as well as in the
canonical case. We also show that such a non-monotonous temperature
dependence can naturally lead to a $\phi_0/2$ periodicity of the
persistent currents, where $\phi_0$=h/e. There is a crossover
temperature $T^*$, below which persistent currents increase in
amplitude with temperature while they decrease above this
temperature. This is in contrast to persistent currents in
rings being monotonously affected by temperature.
$T^*$ is parameter-dependent but of the order of
$\Delta_u/\pi^2k_B$, where $\Delta_u$ is the level spacing of the
isolated ring. For the grand-canonical case $T^*$ is half of that for 
the canonical case.
\end{abstract}
\narrowtext

\section{Introduction}

Although the magnitude of persistent current amplitudes  in
metallic  and semiconductor mesoscopic rings \cite{but}  has
received experimental attention \cite{exp}, much attention has
not been given to qualitative features of the persistent current.
Qualitative features reflect the underlying
phenomena, and are more important than the order of magnitude.
Incidentally, the order of magnitude and sign of the persistent
currents in metallic rings is still not understood. 

With this background in mind, we study the temperature dependence
of persistent currents in a ring strongly coupled to a stub
\cite{buet}. We predict a
non-monotonous temperature dependence of the amplitude of
persistent currents in this geometry both for the grand-canonical as well
as for the canonical case. We show that there is a
crossover temperature ($T^*$) above which it decreases with
temperature and below which it increases with temperature, and
energy scales determining this crossover temperature are
quantified. This is in contrast to the fact that in the
ring, temperature monotonously affects the amplitude of
persistent currents. However, so does dephasing and impurity
scattering, which are again directly or indirectly temperature
dependent \cite{but,Cheung}, except perhaps in very restrictive
parameter regimes where it is possible to realize a Luttinger
liquid in the ring in the presence of a potential barrier
\cite{krive}. Recent study, however, shows that in the
framework of a Luttinger liquid, a single potential
barrier leads to a monotonous temperature dependence of the
persistent currents for non-interacting as well as for
interacting electrons \cite{mos99}.
We also show a temperature-induced switch over
from $\phi_0$ periodicity to $\phi_0/2$ periodicity.
This is a very non-trivial temperature dependence
of the fundamental periodicity
that cannot be obtained in the ring geometry.

There is also another motivation behind studying the temperature
dependence of persistent currents in this ring-stub system. In
the ring, the monotonous behavior of the persistent current
amplitude with temperature stems from the fact that the states
in a ring pierced by a magnetic flux
exhibit a strong parity effect \cite{Cheung}. 
There are two ways of defining this parity effect in the
single channel ring (multichannel rings can be generalized using
the same concepts and mentioned  briefly at the end of this
paragraph). In the single-particle picture (possible
only in the absence of electron-electron interaction), it
can be defined as follows: states with an
even number of nodes in the wave function carry diamagnetic
currents (positive slope of the eigenenergy versus flux)
while states with an odd number of nodes in the
wave function carry paramagnetic currents (negative
slope of the eigenenergy versus flux) \cite{Cheung}. In the
many-body picture (without any electron-electron interaction),
it can be defined as follows:
if $N$ is the number of electrons (spinless)
in the ring, the persistent
current carried by the $N$-body state is diamagnetic if $N$ is
odd and paramagnetic if $N$ is even \cite{Cheung}. Leggett
conjectured \cite{leg} that this parity effect remains unchanged
in the presence of electron-electron interaction and impurity
scattering of any form. His arguments can be simplified to say
that when electrons move in the ring, they pick up three
different kinds of phases: 1) the Aharonov-Bohm phase due to the flux
through the ring, 2) the statistical phase due to electrons being
Fermions and 3) the phase due to the wave-like motion of electrons
depending on their wave vector. The parity effect is due to
competition between these three phases  along with the
constraint that the many-body wave function satisfy the periodic
boundary condition (which means if one electron is taken around
the ring with the other electrons fixed, the many-body
wave function should pick up a phase of 2$\pi$ in all).
Electron-electron interaction or simple potential scattering
cannot introduce any additional phase, although it can change the
kinetic energy or the wave vector and hence modify the third
phase. Simple variational calculations showed that the parity
effect still holds \cite{leg}. Multichannel rings can be
understood by treating impurities as perturbations to decoupled
multiple channels,
which means small impurities  just open up
small gaps at level crossings within the Brillouin zone
and keep all qualitative features
of the parity effect unchanged.
Strong impurity scattering in
the multichannel ring can, however, introduce strong level
correlations,  which is an additional phenomenon. Whether and
how the parity effect gets modified by these correlations is an
interesting problem.
   
In a one-dimensional (1D) system where we have a stub of length
$v$ strongly coupled to a ring of length $u$ (see the left bottom
corner in Fig.~1), we can have a bunching of levels with the same
sign of persistent currents, \cite{Deo95} i.e., many consecutive
levels carry persistent currents of the same sign. 
This is essentially a breakdown of the parity effect.
The parity effect breaks down in this single channel
system because there is a new phase that does not
belong to any of the three phases discussed by Leggett
and mentioned in the preceding paragraph. This new
phase cancels the statistical phase and so the N-body
state and the (N+1)-body state behave in similar ways or carry
persistent currents of the same sign \cite{deo96,sre}. When the
Fermi energy is above the value where we have a
node at the foot of the stub (that results in a transmission
zero in transport across the stub), there is an additional phase
of $\pi$ arising due to a slip in the Bloch phase \cite{deo96}
(the Bloch phase is the third kind of phase discussed above, but the
extra phase $\pi$ due to slips in the Bloch phase is completely
different from any of the three phases discussed above because
this phase change of the wave function is not associated with a
change in the group velocity or kinetic energy or the wave
vector of the electron \cite{deo96,sre}). 
The origin of this phase slip can be understood by
studying the scattering properties of the stub structure.
One can map the stub into a $\delta$-function potential
of the form $k \cot (kv) \delta (x-x_0)$ \cite{deo96}. 
So one can see that
the strength of the effective potential is $k \cot (kv)$ and is energy
dependent. Also the strength of the effective potential is
discontinuous at $kv=n \pi$. Infinitesimally above $\pi$ an electron
faces a positive potential while infinitesimally below it faces
a negative potential. As the effective potential is discontinuous
as a function of energy, the scattering phase, which is otherwise
a continuous function of energy, in this case turns out to be
discontinuous as the Fermi energy sweeps across the point
$kv=\pi$. As the scattering phase of the stub is discontinuous,
the Bloch phase of the electron in the ring-stub
system is also discontinuous.
This is pictorially demonstrated in Figs.~2 and 3 of Ref. \cite{deo96}.
In an energy scale
$\Delta_u\propto 1/u$ (typical level spacing for the isolated
ring of length $u$) if there are $n_b\sim\Delta_u/\Delta_v$
(where $\Delta_v\propto 1/v$, the typical level spacing of the
isolated stub) such phase slips, then each phase slip gives rise
to an additional state with the same slope and there are $n_b$
states of the same slope or ithe same parity bunching together with a
phase slip of $\pi$ between each of them \cite{deo96}.  The
fact that there is a phase slip of $\pi$ between two states of
the same parity was generalized later,
arguing from the oscillation theorem, which is
equivalent to Leggett's conjecture for the parity effect
\cite{lee}. Transmission zeros are an inherent property of Fano
resonance generically occurring in mesoscopic systems and this
phase slip is believed to be observed \cite{the} in a
transport measurement \cite{sch}. For an elaborate discussion on
this, see Ref. \cite{tan}. A similar case was studied in Ref.
\cite{wu}, where they show the transmission zeros and abrupt
phase changes arise due to degeneracy of ``dot states'' with
states of the ``complementary part'' and hence these
are also Fano-type resonances.

The purpose of this work is to show a very non-trivial
temperature dependence of persistent currents due to the
breakdown of the parity effect. The
temperature effects predicted here, if observed experimentally,
will further confirm the existence of parity-violating states,
which is a consequence of this new phase. To be precise, 
the new phase is the key source of the results discussed
in this work. 

\section{Theoretical treatment}

We concentrate on the single channel system to bring out the
essential physics. The multichannel ring also shows a very strong
bunching of levels even though the rotational symmetry is
completely broken by the strongly coupled stub and wide gaps
open up at the level crossings \cite{sre} within the
Brillouin zone. Hence let us consider
a one-dimensional loop of circumference $u$  with a
one-dimensional stub of length $v$, which contain  noninteracting
spinless electrons. The quantum-mechanical potential is zero
everywhere. A magnetic flux $\phi$ penetrates  the ring (see the
left bottom corner in Fig.~1). In this paper we consider both the
grand-canonical case  (when the particle exchange with a
reservoir at temperature $T$ is present and the reservoir fixes
the chemical potential $\mu$; in this case we will denote the
persistent current as $I_\mu$) and the canonical case (when the
number $N$ of particles in the ring-stub system is conserved; in
this case we will denote the persistent current as $I_N$).  For
the grand canonical  case we suppose that the coupling to a
reservoir is weak enough  and the eigenvalues of electron wave
number $k$ are not affected by the reservoir \cite{Cheung}. They
are defined by the following equation \cite{Deo95}. 
   
   \begin{equation}
    \cos(\alpha)=0.5\sin(ku)\cot(kv)+\cos(ku),
    \label{Eq1}
   \end{equation} 
   \\   
   \noindent 
where $\alpha=2\pi\phi/\phi_0$, with $\phi_0=h/e$ being the flux
quantum. Note that Eq. (\ref{Eq1}) is obtained under the
Griffith boundary conditions, \cite{Griffith} which take into
account both the continuity  of an electron wave function and
the conservation of current at  the junction of the ring
and the stub; and the hard-wall boundary condition at the dead end of
the stub. Each of the roots $k_n$ of Eq.(\ref{Eq1}) determines
the one-electron   eigenstate with an energy
$\epsilon_n=\hbar^2k_n^2/(2m)$ as a function of  the magnetic
flux $\phi$. Further we calculate the persistent current 
$I_{N/\mu}=-\partial F_{N/\mu}/\partial \phi$  \cite{Byers},
where $F_N$ is  the free energy for the regime $N=const$ and
$F_\mu$ is the thermodynamic  potential for the regime
$\mu=const$. In the latter case for the system  of
noninteracting electrons the problem is greatly simplified as we
can use the Fermi distribution function 
$f_0(\epsilon)=(1+\exp[(\epsilon-\mu)/T] )^{-1}$ when we fill up
the  energy levels in the ring-stub system and we can write the 
persistent current as follows \cite{Cheung}.
   
   \begin{equation}
    I_\mu=\sum_n I_n f_0(\epsilon_n),
    \label{Eq2}
   \end{equation} 
   \\
   \noindent 
where $I_n$  is a quantum-mechanical current carried by the 
$n$th level and is given by \cite{Deo95}
   
   \begin{equation}
    \frac{\hbar I_n}{e}=\frac{2k_n\sin(\alpha)}{\frac{u}{2}\cos(k_nu)
   \cot(k_nv)-[\frac{v}{2}{\rm cosec}^2(k_nv)+u]\sin(k_nu)}.
    \label{Eq3}
   \end{equation} 
   \\
   For the case of $N=const$ we must calculate the partition
   function $Z$, which determines the free energy  $F_N=-T\ln(Z)$
   \cite{Landau},
   
   \begin{equation}
    Z=\sum_m \exp\left( -\frac{E_m}{T} \right),
    \label{Eq4}
   \end{equation} 
   \\
   \noindent 
where $E_m$ is the energy of a many-electron level. For the 
system of $N$ spinless noninteracting electrons $E_m$ is a sum
over $N$  different (pursuant to the Pauli principle)
one-electron energies  $E_m=\sum_{i=1}^{N} \epsilon_{n_i}$,
where the index $m$ numbers  the different series
$\{\epsilon_{n_1},...,\epsilon_{n_N}\}_m$. For instance, the 
ground-state energy is $E_0=\sum_{n=1}^{N}\epsilon_n$.

\section{Results and discussions}

First we consider the peculiarities of the persistent current
$I_\mu$, $i.e.,$ for the regime $\mu=const$. In this case the
persistent current is determined by Eqs.(\ref{Eq1})-(\ref{Eq3}).
Our calculations show that the character of the temperature
dependence of the persistent currents is essentially dependent
on the position of the Fermi level $\mu$ relative to the
groups of levels with similar currents.  If the
Fermi level lies symmetrically between two groups (which occurs
if $u/\lambda_F=n$ or $n+0.5$, where $n$ is an integer and
$\lambda_F$ is the Fermi wavelength), then the current changes
monotonously with the temperature that is depicted in Fig.~1
(the dashed curve). In this case the low-lying excited levels
carry a current which is opposite to that of the ground-state;
the line shape of the curve is similar to that of the
ring \cite{Cheung}. On the other hand, if the Fermi level lies
within a group ($u/\lambda_F\sim n+0.25$) then low-lying
excited states carry persistent currents with the same sign. In
that case there is an increase of a current at low temperatures
as shown in Fig.~1 (the dotted curve).  At low temperatures the
currents carried by the low-lying excited states add up with the
ground-state current. However, these excited states are only
populated at the cost of the ground state population. 
Although in the clean ring higher levels carry larger
persistent currents, this is not true for the ring-stub system.
This is because the scattering properties of the stub are
energy-dependent and at a higher energy the stub can
scatter more strongly. Hence a lot of energy scales such as
temperature, Fermi energy and number of levels populated
compete with each other to determine the temperature
dependence. A considerable
amount of enhancement in persistent current amplitudes as
obtained in our calculations appears for all choices of
parameters whenever the Fermi energy is approximately
at the middle of a group of levels that have the same
slope. At
higher temperatures when a large number of states get populated,
the current  decreases exponentially. So in this case the
current amplitude has a maximum as a function of the temperature
and we can define the temperature corresponding to the maximum
as the crossover temperature $T^*$.

It is worth mentioning that in the ring system, although there is
no enhancement of persistent currents due to temperature, one
can define a crossover temperature below which persistent
currents decrease less rapidly with temperature. Essentially
this is because at low temperatures thermal excitations are not
possible because of the large single-particle level spacings.
Hence this crossover temperature is the same as the energy scale
that separates two single-particle levels, $i.e.,$ the crossover
temperature is proportional to the  level spacing
$\Delta=hv_F/L$ in the ideal ring at the Fermi surface, where
$v_F$ is the Fermi velocity and $L$ is the length of the ring. 
The crossover temperature obtained by us in the ring-stub system
is of the same order of magnitude, $i.e.,$  $\Delta_u=hv_F/u$,
although different in meaning.

In the case of $u/\lambda_F=n+0.25$ at low temperatures we show
the possibility of obtaining $\phi_0/2$ periodicity, although
the parity effect is absent in this system. This is shown in Fig.~2,
where we plot $I_{\mu}/I_0$ 
versus  $\phi/\phi_0$ at a temperature
$k_BT/\Delta_u$=0.01  in solid lines, which clearly show a
$\phi_0/2$ periodicity. Previously two mechanisms are known that
can give rise to $\phi_0/2$ periodicity of persistent currents.
The first is due to the parity effect \cite{los}, which does not
exist in our system, and the second
is due to the destructive interference of the
first harmonic that can only appear in a system coupled to a
reservoir so that the Fermi energy is an externally adjustable
parameter. The later mechanism can be understood by putting
$k_FL$=$(2n\pi+\pi /2)$ in eq. 2.11 in Ref. \cite{Cheung}. If
this later case is the case in our 
situation, then the periodicity should remain
unaffected by temperature and for fixed $N$ we should only get
$\phi_0$ periodicity \cite{Cheung}, because then the
Fermi energy is not
an externally adjustable parameter but is determined by $N$. We
show in Fig.~2 (dashed curve) 
that the periodicity changes with temperature and
in the next two paragraphs 
we will also show that one can obtain $\phi_0/2$
periodicity for fixed $N$. The dashed curve in Fig.~2 is
obtained at a temperature $k_BT/\Delta_u$=0.15  and it shows a
$\phi_0$ periodicity. As it is known,
the crossover temperature depends on  the harmonic number $m$:
$T^*_m=T^*/m$ \cite{Cheung}, in this case a particular harmonic
can actually increase with temperature initially and decrease
later, different harmonics reaching  their peaks at different
temperatures. Therefore, the second harmonic that peaks at a
lower temperature than the first harmonic can exceed the first
harmonic in certain temperature regimes. At higher temperatures
it decreases with the temperature faster than the first harmonic
and so at higher temperature $\phi_0$ periodicity is
recovered.  

In view of a strong dependence of the considered features on
the  chemical potential, we consider further the persistent
current $I_N$ in the ring-stub system with a fixed number of
particles $N=const$. In this  case we calculate the persistent
current using the partition function  (Eq.(\ref{Eq4})). 

The numerical calculations show that in this case
there is also a non-monotonous temperature dependence of the
persistent current  amplitude in the canonical case as in the 
grand-canonical case. 
This is shown in Fig.~1 by the solid curve. The
maximum of $I_N(T)$ is more pronounced if $v/u$ is
large and the number of electrons ($N$) is small. Besides, if the
number of electrons is more than $n_b/2$, then the maximum does
not exist. The crossover temperature is higher by a factor 2
as compared to that in $I_\mu$. This was also found for the 1D
ring \cite{Cheung,Loss92}, where, as mentioned before, the
crossover temperature has a different meaning. To show that one
can have $\phi_0/2$ periodicity for fixed $N$, we plot in the
inset of Fig.~2 the first harmonic $I_1/I_0$ (solid curve) and the
second harmonic $I_2/I_0$ (dotted curve) of $I_N$
for $N$=5, $v$=7$k_F$ and
$u$=2.5$k_F$. At low temperature the second harmonic exceeds
the first harmonic because the stub reduces the level spacing and in
a sense can adjust the Fermi energy in the ring to create
partial but not exact destruction of the first harmonic.
There are distinct temperature regimes where $I_1$
exceeds $I_2$ and vice versa and the two curves peak in
completely different temperatures. $I_2$ also exhibits more than
one maxima.
Experimentally different harmonics can be measured separately
and the first harmonic as shown in Fig.~2 can show tremendous
enhancement with temperature.

An important conclusion that can be made from Fig.~2 is that
observation of $\phi_0/2$ periodicity as well as $\phi_0$ is
possible even in the absence of the parity effect quite naturally
because the absence of the parity effect also means one can obtain
an enhancement of the persistent current amplitude with temperature,
and as a result an enhancement of a particular harmonic with
temperature, resulting in different harmonics peaking at
different temperatures.

\section{Conclusions}

In summary, we would like to state that the temperature dependence of
persistent currents in a ring strongly coupled to a stub exhibits
very nontrivial features. Namely, at small temperatures it can
show an enhancement of the amplitude of persistent currents
in the grand-canonical as well as in the canonical case.
The fundamental periodicity of the persistent currents can change
with temperature.
If detected experimentally, these can lead to a better
understanding of the qualitative features of persistent
currents. It will also confirm the existence of parity-violating
states that is only possible if there is a new phase apart from
the three phases considered by Leggett \cite{leg} while
generalizing the parity effect. 
This new phase is the sole cause of the nontrivial temperature
dependence. There is a crossover temperature
$T^*$ above which the amplitude of persistent currents decreases
with temperature. How the crossover temperature is affected by
electron correlation effects and dephasing should lead to
interesting theoretical and experimental explorations in the future.

Finally, with the large discrepancies between theory and
experiments for the persistent currents in disordered rings,
one cannot completely rule out
the possibility of parity violation in the ring system as well.
The stub is not the only way to produce this new phase
that leads to a violation of the parity effect. There can be more
general ways of getting transmission zeros \cite{lee} that may
also be parity violation. In that case, the ring-stub system may
prove useful as a theoretical model to understand  the
consequences of parity violation. Its consequences on the
temperature dependence shown here may motivate future works in
this direction.

   %----------------------------------------------------------------------

    %---------------------------------------------------------------
   
      \centerline{\Large {\bf Figure captions} }
      \ \\
   
Fig.~1. The ring of length $u$ with a stub (resonant cavity) of
length $v$ threaded by a magnetic flux $\phi$ (left bottom
corner). The dependence of the current amplitude $I_\mu$ in units
of  $I_0=ev_F/u$ on the temperature $T$ in units of
$\Delta_u/2\pi^2k_B$  for the regime $\mu=const$  with
$v=15\lambda_F$ and $u=(5+x)\lambda_F$ at $x=0$ (dashed curve)
and $x=0.25$ (dotted curve); and $I_N/I_0$ for the isolated ring-stub
system with $v/u=10$, and $N=3$  (solid curve). For the
appropriate scale the curves 2 and 3 are multiplied by  factors 
of 3 and 15, respectively.

Fig.~2. The dependence of the persistent current $I_\mu$ in units
of  $I_0=ev_F/u$ on the magnetic flux $\phi$ in units of
$\phi_0$ for the regime  $\mu=const$ with $v=15\lambda_F$ and
$u=5.25\lambda_F$ for $T/\Delta_u=0.01$ (dashed curve) and
$T/\Delta_u=0.15$ (solid curve).  The curve 2 is multiplied by a
factor of 5 for the appropriate scale. The inset shows the first
harmonic $I_1$ (solid curve) and second harmonic $I_2$ (dotted
curve) of $I_N$ in units of $I_0$ for N fixed at 5, $v$=7$k_F$ and
$u$=2.5$k_F$ versus temperature in units of
$\Delta_u/2\pi^2k_B$.
   
   %---------------------------------------------------------------------
   

\begin{thebibliography}{99}
\bibitem[*]{eml}email: deo@boson.bose.res.in   
\bibitem{but} M. B{\"u}ttiker, Y. Imry, and R. Landauer,
Phys. Lett. {\bf 96A}, 365 (1983).
\bibitem{exp} L. P. Levy et al, Phys. Rev. Lett. {\bf 64},
2074 (1990); V. Chandrasekhar et al, Phys. Rev. Lett. {\bf 67},
3578 (1991); D. Mailly et al, Phys. Rev. Lett. {\bf 70}, 2020 (1993).
\bibitem{buet} The weak coupling limit was earlier studied by
M. B{\"u}ttiker, Phys. Scripta T {\bf 54}, 104 (1994). Coupled rings
was studied by T.P. Pareek and A.M. Jayannavar, Phys. Rev. B 
{\bf 54}, 6376 (1996).
\bibitem{Cheung} H.F.Cheung, Y.Gefen, E.K.Riedel, W.H.Shih,
Phys. Rev. B {\bf 37}, 6050 (1988).
\bibitem{krive} I.V. Krive et al, Phys. Rev. B {\bf 52}, 16451 (1995);
A.S.Rozhavsky, J. Phys.: Condens. Matter {\bf 9}, 1521 (1997);
I. V. Krive et al, cond-mat/9704151.
\bibitem{mos99} M.V. Moskalets, Physica E {\bf 5}, 124 (1999).
\bibitem{leg} A.J. Leggett in: Granular nano-electronics,
eds. D. K. Ferry, J.R. Barker and C. Jacobony, NATO ASI Ser.
B {\bf 251} (Plenum, New York, 1991) p. 297.
\bibitem{Deo95} P.Singha Deo, Phys. Rev. B {\bf 51}, 5441 (1995).
\bibitem{deo96} P. Singha Deo, Phys. Rev. B {\bf 53}, 15447 (1996).
\bibitem{sre} P. A. Sreeram and P. Singha Deo, Physica B {\bf 228},
345 (1996).
\bibitem{lee} H.-W.Lee, Phys. Rev. Lett., {\bf 82}, 2358 (1999).
\bibitem{the} P.Singha Deo and A.M.Jayannavar, Mod. Phys. Lett. B {\bf 10}, 
787 (1996);
P.Singha Deo, Solid St. Communication {\bf 107}, 69 (1998);
C.M.Ryu et al, Phys. Rev. B {\bf 58}, 3572 (1998);
Hongki Xu et al, Phys. Rev. B, {\bf 57}, 11903 (1998).
\bibitem{sch} R. Schuster et al, Nature {\bf 385}, 417 (1997).
\bibitem{tan} T. Taniguchi and M. B\"uttiker, Phys. Rev. B {\bf 60},
13814 (1999).
\bibitem{wu} J.Wu et al, Phys. Rev. Lett. {\bf 80}, 1952 (1998).
\bibitem{Griffith} S. Griffith, Trans. Faraday. Soc. {\bf 49}, 650 (1953).
\bibitem{Byers} N. Byers, C.N. Yang, Phys. Rev. Lett. {\bf 7}, 46 (1961);
F. Bloch, Phys. Rev. B {\bf 2}, 109 (1970).
\bibitem{Landau} L.D. Landau, E.M. Lifschitz, (1959) Statistical Physics 
(Pergamon, London). 
\bibitem{los} D. Loss and P. Goldbart, Phys. Rev. B {\bf 43}, 13762
(1991).
\bibitem{Loss92} D. Loss, Phys. Rev. Lett. {\bf 69}, 343 (1992).
   
    \end{thebibliography}
      
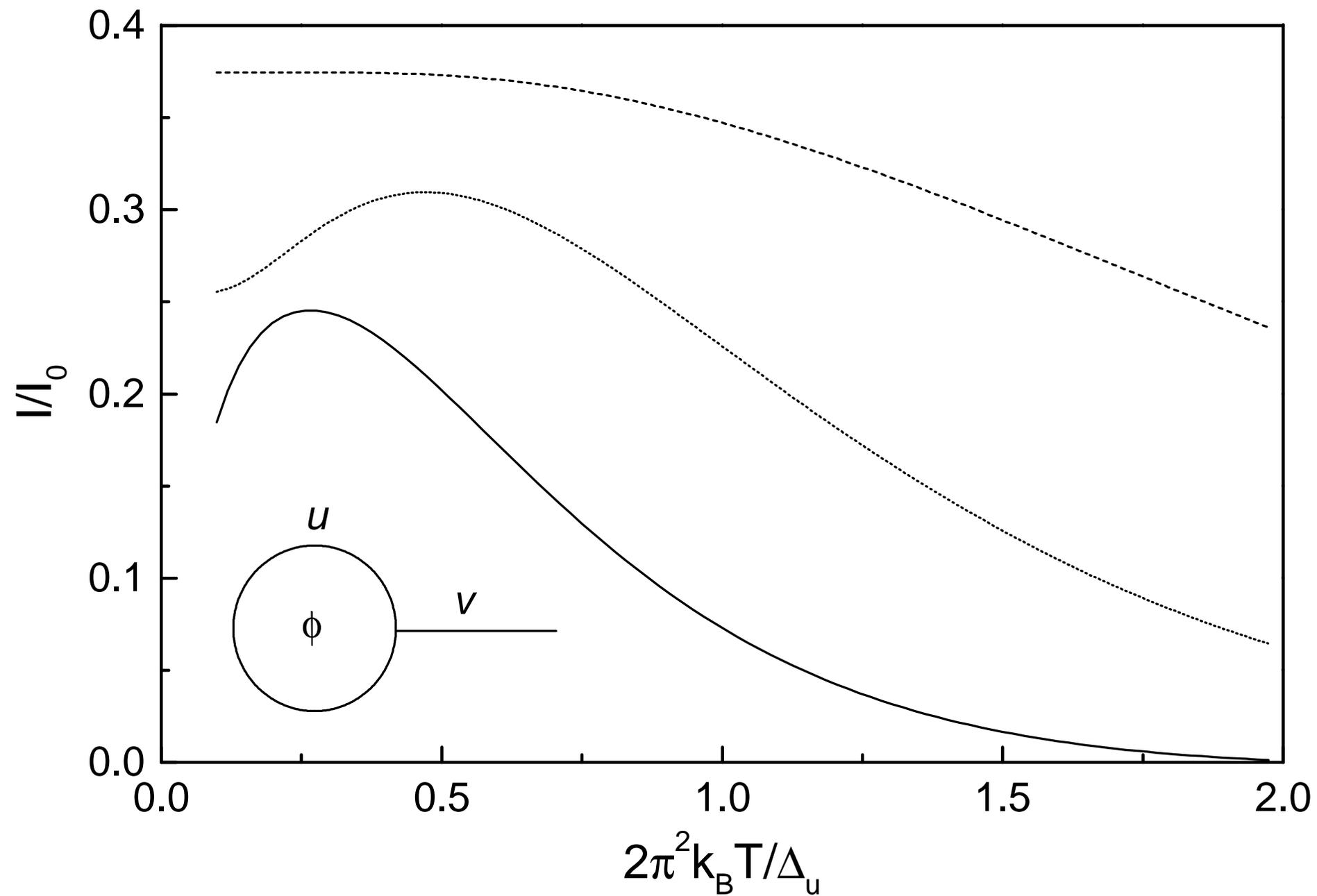
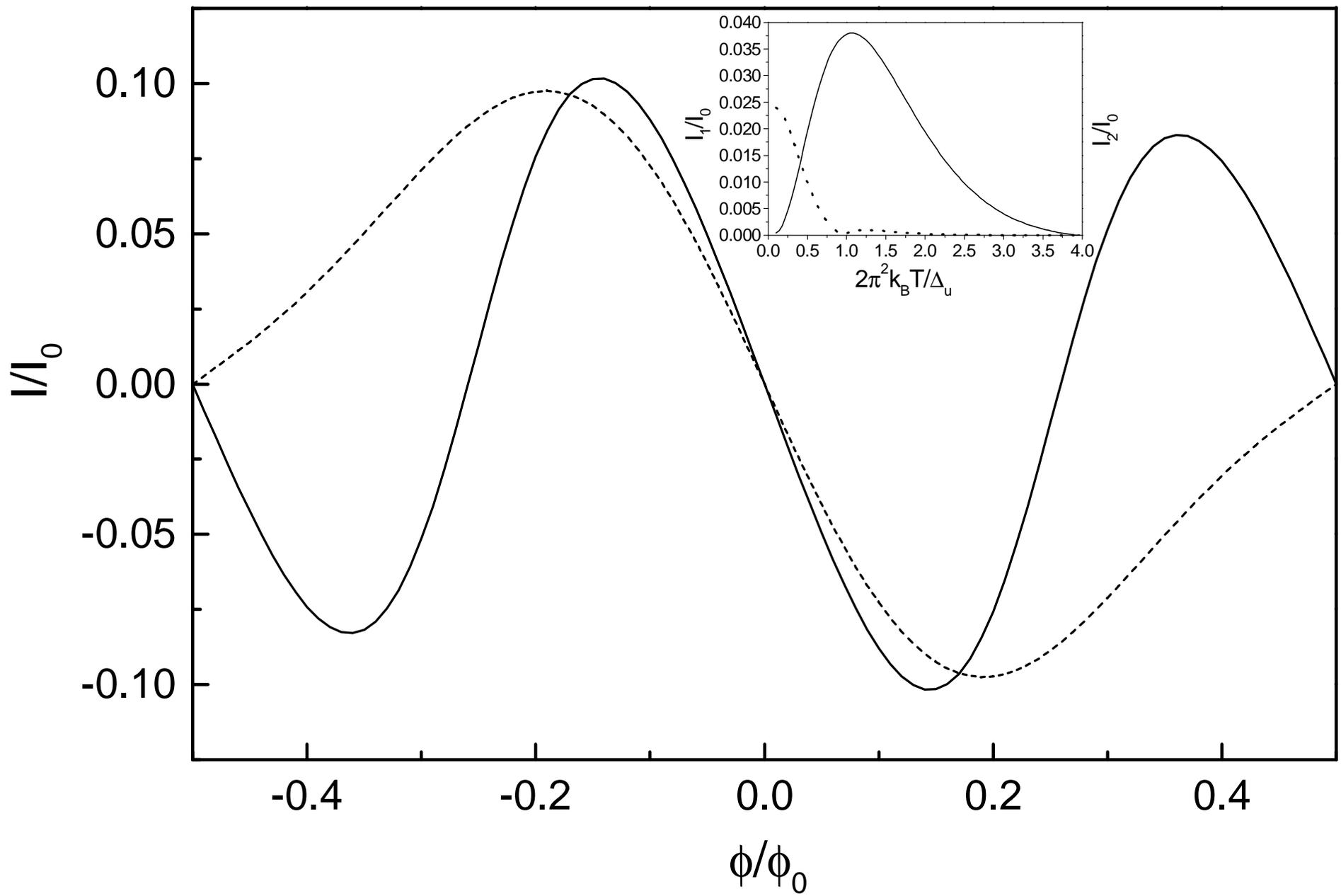
\end{document}